# Discussion about Attacks and Defenses for Fair and Robust Recommendation System Design*

Mirae Kim
Department of Artificial Intelligence
Sungkyunkwan University
Suwon, Korea
miraekim1992@g.skku.edu

Simon Woo
Department of Artificial Intelligence
Sungkyunkwan University
Suwon, Korea
swoo@g.skku.edu

**ABSTRACT**

Information has exploded on the Internet and mobile with the advent of the big data era. In particular, recommendation systems are widely used to help consumers who struggle to select the best products among such a large amount of information. However, recommendation systems are vulnerable to malicious user biases, such as fake reviews to promote or demote specific products and attacks that steal personal information. Such biases and attacks compromise the fairness of the recommendation model and infringe the privacy of users and systems by distorting data. Recently, deep-learning collaborative filtering recommendation systems have shown to be more vulnerable to this bias. In this position paper, we examine the effects of bias that cause various ethical and social issues and discuss the need for designing a robust recommendation system for fairness and stability.

**ACM Reference format:**

## 1 Introduction

The amount of data on the Internet grows exponentially every year as the use of web content increases. Consumers have difficulty choosing the information and products they want from these excessive choices. The recommendation systems provide the best products and information consumers might need by algorithmically mining and presenting valuable data[1].

However, recently recommendation systems have been known to be vulnerable to malicious attacks[2]. In particular, such recommendation systems may be biased towards fake reviews to promote or demote certain items. In addition, attacks that steal personal information from recommendation systems can cause problems such as privacy[3, 4] and distortion of user opinions. Therefore, this undermines the fairness and security of the entire recommendation ecosystem.

This position paper presents popular attack and defense techniques for recommendation systems. We show that much research concerning attack and defense is necessary to provide fairness and security in the recommendation models. Through this position paper, we discuss the impact of such attacks and strategies for designing a robust recommendation model.

## 2 Attacks & Defenses

This section describes the widespread attack and defense techniques for the recommendation model. We explain and investigate several attacks on recommendation systems and their impact on the fairness and stability of the model. We also explore how to design a secure and robust recommendation system against these attacks.

### 2.1 Attacks

**Shilling Attack.** A typical attack can poison the input data[5], leading the model to make incorrect recommendations. The traditional collaborative filtering[6] model uses a statistically designed shilling attack. In particular, random attack[7], average attack[8], bandwagon attack[9], and segment attack[10] are commonly used algorithms. However, these statistical-based attack algorithms are unsuitable for the recently used deep learning recommendation models. Adversarial attacks are being researched as attacks against deep learning-based recommendation models[11].

**Table 1 Shilling Attack Methods**

| Random Attack | Random ratings for all items |
|---|---|
| Average Attack | Average rating of individual items |
| Bandwagon Attack | Weighted ratings for popular items |
| Segment Attack | Weighted ratings for same target group |

**Adversarial Attack.** Adversarial attack[12] changes the model's output by injecting a user profile with noise into the existing inputs. Several studies[13-16] have shown that deep learning models are more vulnerable to adversarial attacks than shilling attacks. Representative adversarial attack algorithms are FGSM[13] and C&W[14]. Since these methods are originally used in image classification models, research is underway to optimize the attack vectors according to the characteristics of the recommendation models whose inputs are interdependent. In addition, generative adversarial network (GAN)[16]-based adversarial attacks[15] are also being investigated, where an attacker creates an undetectable user profile. Research is also being conducted to build optimal recommendation models with various inputs.

In summary, recommendation systems can be easily exposed to various attacks, as shown in Fig. 1. In particular, the need for research on adversarial attacks is increasing as the vulnerability of the deep learning-based models to adversarial attacks is being reported.

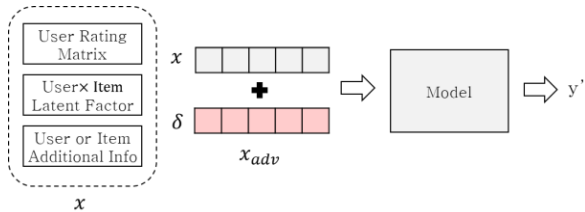

**Figure 1 Adversarial Attack on Recommendation System**

## 2.2 Effects of Attacks

Malicious attacks can easily degrade the performance of a recommendation system. Such attacks can undermine not only the accuracy of the model but also the fairness and stability, causing several ethical and social problems in the entire recommendation ecosystem. This section examines the effects and problems of such attacks on the recommendation model.

**Fairness.** The fairness of the recommended information is compromised when the recommendation system is attacked. In particular, an attacker can increase, decrease, and manipulate the popularity of desired target product. Zhuoran Liu et al. [17] showed that sellers could promote the ranking of certain items by creating fake profiles to promote their products, and this can weaken the fairness of recommendation systems.

The attack can lead the model itself to make incorrect recommendations. It can be used to diminish the customer lock-in effect by reducing the accuracy performance of the recommendation model. He et al. [18] showed that the model had a 13x higher performance degradation to adversarial attacks than shilling attacks. This result means that adversarial attacks are likely to prevent consumers from feeling the reliability, fairness, and trust of recommended products.

**Security.** A malicious attacker can infer a user's personal information through the recommendation[3, 4]. Existing works suggest that if the attacker had access to the system's output and secondary information, it could extract the user's entire history. Moreover, attacks can distort users' opinions[25] and cause biases, and attacks can further infer user data by tracing back model parameters.

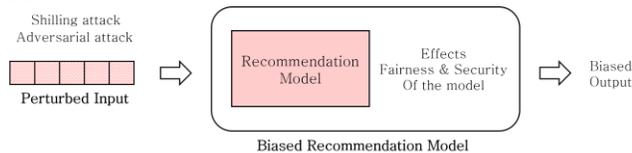

**Figure 2 Effects of Attacks on Recommendation Systen**

Because of the model's vulnerability, the attacks can access and change sensitive user data. Therefore, it is necessary to design a robust model to prevent easy access to personal information in the recommendation system.

## 2.3 Defenses

Defense techniques against the attacks mentioned above include adversarial training[20], and knowledge distillation[21], attack detection[19]. While adversarial training is mainly used, there is a limitation where it is only suitable for a model using a specific algorithm or input. Therefore, there is a need for research on how to design a recommendation model that works effectively for various attack algorithms.

**Adversarial Training.** It is a method using adversarial samples in the training process[12]. Even if a malicious profile is injected, it is possible to create a recommendation model that generally reacts to the attacks because the fake profile has already been used in the training procedure. A representative study is the AMR method[22]. J. Tang et al. showed that the negative impact of adversarial attacks measured by nDCG decreases from -8.7% to -1.4% when using adversarial training over classical training.

**Knowledge Distillation.** Knowledge distillation[21] is the process of transferring knowledge from a large model (teacher) to a smaller one (student). Moreover, the student model is known to be robust against adversarial attacks. Yali et al. [23] investigate the defense technique using distillation in a recommendation system. They improved the system's robustness with the concept of stage-wise hints training and randomness. The defense method can reduce the success rate of malicious user attacks and maintain prediction accuracy similar to standard neural recommendation systems. However, knowledge distillation is not widely used as a defense technique and has rarely been studied.

**Attack Detection.** Attack detection[19] is an approach that detects fake input generated by attackers. Current research has limitations due to the low generality of the algorithm, lack of optimization design model, and difficulty in selecting user profile attributes. In particular, Wei et al. [24] proposed a shilling attack detection structure based on the rating time-series analysis of anomalous group users, where 90% of attack detection performance is achieved with 100 attack sizes. It is necessary to examine how to design a general detection module that can detect various attacks in the future.

## 3 Conclusion

Recommendation systems are exposed to a variety of attacks and can easily degrade performance. In particular, deep learning models are vulnerable to adversarial attacks, resulting in fairness and security problems. In this position paper, we investigate various types of attack and defense techniques for recommendation systems. Through this position paper, we hope that future work will continue exploring new ways to build a robust recommendation system against different attacks and maintain the fairness and privacy of the entire end-to-end recommendation ecosystem.

## ACKNOWLEDGMENTS

This work was supported by Institute of Information & communications Technology Planning & Evaluation (IITP) grant funded by the Korea government (MSIT) (No.2019-0-00421, Artificial Intelligence Graduate School Program (Sungkyunkwan University))